\documentclass{article}
\usepackage[utf8]{inputenc}
\usepackage{authblk}
\usepackage{setspace}
\usepackage[margin=1.25in]{geometry}
\usepackage{graphicx}
\graphicspath{ {./figures/} }
\usepackage{amsmath}
\usepackage{lineno}
\usepackage{bm}
\usepackage{bigints}

\usepackage{subfigure}
\usepackage{cite}

\usepackage{color}

\providecommand\JournalTitle[1]{#1}

\title{Extrapolated speckle-correlation imaging}

\author[1]{Yuto Endo}
\author[2]{Jun Tanida}
\author[1]{Makoto Naruse}
\author[1*]{Ryoichi Horisaki}

\affil[1]{Department of Information Physics and Computing, Graduate School of Information Science and Technology, the University of Tokyo, 7-3-1 Hongo, Bunkyo-ku, Tokyo 113-8656, Japan}
\affil[2]{Department of Information and Physical Sciences, Graduate School of Information Science and Technology, Osaka University, 1-5 Yamadaoka, Suita, Osaka 565-0871, Japan}
\affil[*]{Corresponding author. Email: horisaki@g.ecc.u-tokyo.ac.jp}

\date{}

\begin{document}

\maketitle

\begin{abstract}
Imaging through scattering media is a longstanding issue in a wide range of applications, including biomedicine, security, and astronomy.
Speckle-correlation imaging is promising for non-invasively seeing through scattering media by assuming shift-invariance of the scattering process called the memory effect.
However, the memory effect is known to be severely limited when the medium is thick.
Under such a scattering condition, speckle-correlation imaging is not practical because the correlation of the speckle decays, reducing the field of view.
To address this problem, we present a method for expanding the field of view of single-shot speckle-correlation imaging by extrapolating the correlation with a limited memory effect.
We derive the imaging model under this scattering condition and its inversion for reconstructing the object.
Our method simultaneously estimates both the object and the decay of the speckle correlation based on the gradient descent method.
We numerically and experimentally demonstrate the proposed method by reconstructing point sources behind scattering media with a limited memory effect.
In the demonstrations, our speckle-correlation imaging method with a minimal lensless optical setup realized a larger field of view compared with the conventional one. 
This study will make techniques for imaging through scattering media more practical in various fields.
\end{abstract}

\section*{Keywords}
Computational imaging;
Imaging through scattering media;
Memory effect;
Phase retrieval;
Speckle correlation


\section{Introduction}
Seeing through scattering media is an important research topic in optics and photonics because of its wide range of applications.
For example, microscope imaging through biological tissues and telescopic observation through atmospheric turbulence are longstanding issues in biomedicine and astronomy, respectively~\cite{Ntziachristos2010,Ji2017, Davies2012,Watnik2018}.
Recent advancements in optics and information science have driven studies for imaging through strongly scattering media where conventional methods assuming ballistic photons are difficult to apply~\cite{Mosk2012,Horstmeyer2015,Yoon2020}.

Wavefront shaping based on feedback and object retrieval with a transmission matrix are established approaches for imaging through scattering media.
In the wavefront shaping approach, a scattering pattern on an image sensor inside the scattering medium is fed back to a spatial light modulator outside the medium to focus light on the sensor~\cite{Vellekoop2007, Vellekoop2010a, Katz2011}.
In the transmission matrix approach, the scattering process is described as a matrix, and the object is recovered from a single captured image with the inversion of the matrix~\cite{Popoff2010a, Liutkus2014, Horisaki2016}.
The issues with those approaches are invasiveness and complex optical setups for the feedback to optimize the wavefront or the observation of the transmission matrix.

Speckle-correlation imaging is a promising non-invasive approach with a minimal optical setup to address the above issues~\cite{Bertolotti2012,Katz2014a}.
This approach assumes shift-invariance of the scattering impulse response called the memory effect, and the object is recovered from the autocorrelation of the captured speckle pattern by using a phase retrieval algorithm~\cite{Fienup1982,Feng1988,Issac1988,Fienup2013}.
It is also extendable to multidimensional imaging while maintaining its optical simplicity~\cite{Okamoto2019,Horisaki2019,Ehira2021}.

An issue with speckle-correlation imaging is the small field of view due to a limited range of the memory effect when the scattering medium is thick.
To overcome this difficulty, ptychographic methods have been introduced, although these require multi-shot measurements and additional hardware components~\cite{Li2019a,Rosenfeld2021}.
Other interesting methods for single-shot imaging include decomposition of multiplexed speckle correlations and localization of the speckle correlation~\cite{Wang2019,Alterman2021}.
These methods require some specific optical conditions, such as isolated objects or a near-field setting.

Here we propose and demonstrate a method for extending the field of view of single-shot speckle-correlation imaging to address the above issues.
The proposed method takes account of the decay of the speckle correlation under a limited memory effect and extrapolates the correlation in the reconstruction process.
Our method is readily applicable to conventional speckle-correlation imaging methods without any optical modifications.
Therefore, this study will contribute to various imaging applications where scattering causes limitations.

\section{Method}

\begin{figure}[t]
	\centering
	\includegraphics[scale=1]{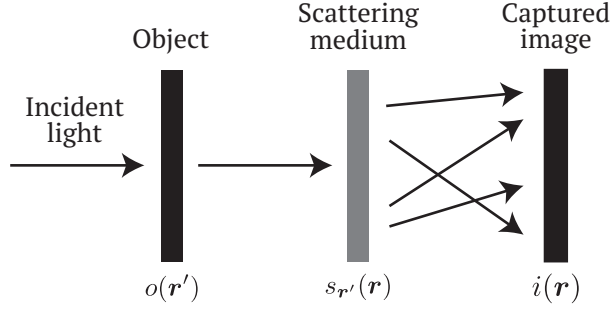}
	\caption{Optical model in the proposed method.}
  \label{fig_method}
\end{figure}

In our method, an object~$o$ is observed as a captured image~$i$ on an image sensor through a scattering process without any additional device, as shown in Fig.~\ref{fig_method}.
This measurement process is written with a shift-variant point spread function~(PSF)~$s$ as
\begin{equation}
i(\bm{r}) = \int o(\bm{r}')s_{\bm{r}'}(\bm{r}) \mathrm{d}^2\bm{r}',
  \label{eq_method_capturedimg}
\end{equation}
where $\bm{r}$ and $\bm{r}'$ are two-dimensional coordinates on the sensor and object planes, respectively.
Here, $s_{\bm{r}'}(\bm{r})$ is the response at $\bm{r}$ on the sensor plane from the impulse at $\bm{r}'$ on the object plane.

In speckle-correlation imaging, we calculate the autocorrelation of the captured image~$i(\bm{r})$ as follows~\cite{Bertolotti2012,Katz2014a}:
\begin{equation}
\begin{aligned}
\relax[i\star i](\bm{r})
&=\int i(\widetilde{\bm{r}})i(\widetilde{\bm{r}}+\bm{r}) \ \mathrm{d}^2\widetilde{\bm{r}} \\
&=\iint o(\bm{r}_1')s_{\bm{r}_1'}(\widetilde{\bm{r}})\ \mathrm{d}^2\bm{r}_1' \int o(\bm{r}_2')s_{\bm{r}_2'}(\widetilde{\bm{r}}+\bm{r})\ \mathrm{d}^2\bm{r}_2' \ \mathrm{d}^2\widetilde{\bm{r}} \\
&=\iint o(\bm{r}_1')o(\bm{r}_2') \int s_{\bm{r}_1'}(\widetilde{\bm{r}})s_{\bm{r}_2'}(\widetilde{\bm{r}}+\bm{r}) \ \mathrm{d}^2\widetilde{\bm{r}}\ \mathrm{d}^2\bm{r}_1'\ \mathrm{d}^2\bm{r}_2'\\
&=\iint o(\bm{r}_1')o(\bm{r}_2') [s_{\bm{r}_1'}\star s_{\bm{r}_2'}](\bm{r})\ \mathrm{d}^2\bm{r}_1'\ \mathrm{d}^2\bm{r}_2'.
\end{aligned}    
  \label{eq_method_forwardmodel_1}
\end{equation}

When the memory effect is limited, $[s_{\bm{r}_1'}\star s_{\bm{r}_2'}](\bm{r})$ in Eq.~(\ref{eq_method_forwardmodel_1}), which is the correlation between the two scattering PSFs from the two impulse positions~$\bm{r}_1'$ and $\bm{r}_2'$, respectively, is described as
\begin{equation}
[s_{\bm{r}_1'}\star s_{\bm{r}_2'}](\bm{r}) \propto \delta(\bm{r}-(\bm{r}_2'-\bm{r}_1')) \ d(\sigma,\bm{r}_2'-\bm{r}_1') + c,
  \label{eq_method_psf_croscor}
\end{equation}
where $c$ is background noise, $d(\sigma,\bm{r})$ is a decay function of the correlation, and $\sigma$ is a parameter for the decay function~\cite{Feng1988,Issac1988,Schot2015}.
By using the paraxial approximation, this decay function is written as
\begin{equation}
d(\sigma,\bm{r}) = \left(\frac{\sigma |\bm{r}|}{\sinh(\sigma |\bm{r}|)}\right)^2.
  \label{eq_method_dacay}
\end{equation}
Here $\sigma=kL/R$ is an unknown in this study, where $k$ is the wave number, $L$ is the thickness of the scattering medium, and $R$ is the distance between the scattering medium and the sensor plane.
A large $\sigma$ represents a limited memory effect caused by a thick scattering medium.

By substituting Eq.~(\ref{eq_method_psf_croscor}) into Eq.~(\ref{eq_method_forwardmodel_1}) and introducing a variable~$\bm{\tau} = \bm{r}_2'-\bm{r}_1'$, the autocorrelation of the captured image~$i(\bm{r})$ is rewritten as
\begin{equation}
\begin{aligned}
\relax[i\star i](\bm{r})&\propto \iint o(\bm{r}_1')o(\bm{r}_2')\bigl(\delta(\bm{r}-(\bm{r}_2'-\bm{r}_1'))\ d(\sigma,\bm{r}_2'-\bm{r}_1') + c\bigr) \mathrm{d}^2\bm{r}_1'\mathrm{d}^2\bm{r}_2' \\
&= \iint o(\bm{r}_1')o(\bm{r}_1'+\bm{\tau})\bigl(\delta(\bm{r}-\bm{\tau}) \ d(\sigma,\bm{\tau}) + c\bigr) \mathrm{d}^2\bm{r}_1'\mathrm{d}^2\bm{\tau} \\
&= \int [o\star o](\bm{\tau})\bigl(\delta(\bm{r}-\bm{\tau}) \ d(\sigma,\bm{\tau}) + c\bigr)\mathrm{d}^2\bm{\tau} \\
&= [o\star o](\bm{r})\ d(\sigma,\bm{r}) +\widetilde{c}\\
&=[\mathcal{A}(o)](\bm{r})\ d(\sigma,\bm{r}) +\widetilde{c},
\end{aligned}    
  \label{eq_method_forwardmodel_2}
\end{equation}
where $\widetilde{c}$ is background noise, and $\mathcal{A}$ denotes an operator of the autocorrelation.
Therefore, by ignoring the background noise, the autocorrelation of the captured image~$i(\bm{r})$ is the product of the autocorrelation of the object~$o(\bm{r}')$ and the decay function~$d(\sigma,\bm{r})$.

In this study, we simultaneously calculate the estimations of both the object~$\widehat{o}$ and the decay parameter~$\widehat{\sigma}$ by solving the following optimization problem
\begin{equation}
\underset{\widehat{o}, \widehat{\sigma}}{\text{arg~min}} ~f(\widehat{o}, \widehat{\sigma}) + g(\widehat{o}),
 \label{eq_method_opt}
\end{equation}
where $f(\widehat{o},\widehat{\sigma})$ is an error function with the forward model in Eq.~(\ref{eq_method_forwardmodel_2}).
The inverse problem of Eq.~(\ref{eq_method_forwardmodel_2}) is ill-conditioned and is difficult to solve by itself, so we introduce a penalty function~$g(\widehat{o})$ for the object to regulate the inversion.
Here we define the error function~$f(\widehat{o},\widehat{\sigma})$ as follows:
\begin{align}
f(\widehat{o},\widehat{\sigma})&= \Bigl\lVert e(\widehat{o},\widehat{\sigma},\bm{r})v(\bm{r})\Bigr\rVert _2^2~,
 \label{eq_method_errorfunc}\\
e(\widehat{o},\widehat{\sigma},\bm{r})&=[\mathcal{A}(\widehat{o}\gamma)](\bm{r}) d(\widehat{\sigma},\bm{r}) - [\mathcal{A}(i)](\bm{r}),
\label{eq_method_notation_e}
\end{align}
where $\gamma(\bm{r}')$ is the support of the object, and $v(\bm{r})$ is the support of the autocorrelation.
Both of the supports are binary patterns.
The object's support constrains the estimated object area~\cite{Fienup1982}.
The autocorrelation's support is introduced to remove the measurement noise on the captured image~$i(\bm{r})$ because the noise concentrates at the central peak of the autocorrelation~$[i\star i](\bm{r})$ and is removed by a central hole on~$v(\bm{r})$.
Here the background noise~$\widetilde{c}$ can be ignored in the forward model in Eq.~(\ref{eq_method_forwardmodel_2}) by computational removal after taking the autocorrelation of the speckle image~$i(\bm{r})$~\cite{Bertolotti2012,Katz2014a, Hofer2018}.

The penalty function~$g(\widehat{o})$ is composed of the following three sub-functions:
\begin{align}
g(\widehat{o})=\alpha_+ g_+(\widehat{o})+ \alpha_{\theta} g_{\theta}(\widehat{o}) + \alpha_{\ell_1} g_{\ell_1}(\widehat{o}),
\label{eq_method_penaltyfunc}
\end{align}
where $g_+(\widehat{o})$, $g_{\theta}(\widehat{o})$, and $g_{\ell_1}(\widehat{o})$ are penalties of the non-negativity, the upper-limitation with the threshold~$\theta$, and the sparsity with the $\ell_1$ norm, respectively.
Here, $\alpha_+$, $\alpha_{\theta}$, and $\alpha_{\ell_1}$ are tuning parameters for each of the sub-functions in the optimization process.
These sub-functions for the object's penalties are defined as
\begin{align}
g_+(\widehat{o})&=\|\min(0,\widehat{o}(\bm{r}'))\|_2^2,
\label{eq_method_nonnega}\\
g_{\theta}(\widehat{o})&=\|\max(\theta,\widehat{o}(\bm{r}'))-\theta\|_2^2,
\label{eq_method_th}\\
g_{\ell_1}(\widehat{o})&=\|\max(0,\widehat{o}(\bm{r}'))\|_1,
\label{eq_method_spa}
\end{align}
where $``\min"$ is an operator outputting the smaller values of the two parenthesized variables, and $``\max"$ is an operator outputting the larger values of the two parenthesized variables, respectively.
Our algorithm in this study supposes sparse objects, since we introduced the sparsity regularization~$g_{\ell_1}$ in Eq.~(\ref{eq_method_spa}), and is therefore not applicable to dense objects.
However, the assumption of the object's sparsity is acceptable in specific applications, such as astronomical observation and super-resolution fluorescence microscopy, where objects are sparse~\cite{Betzig2006, Davies2012, Tehrani2015, Wang2021}.
The limitation given by the sparsity regularization is demonstrated in the numerical analysis, and ways of mitigating it are mentioned in the conclusion.

We solve the optimization problem in Eq.~(\ref{eq_method_opt}) based on an iterative gradient descent algorithm in which the decayed autocorrelation is extrapolated.
The gradient descent steps for each $\widehat{o}$ and $\widehat{\sigma}$ are written as
\begin{align}
\widehat{o}_{k+1}(\bm{r}')&=\widehat{o}_{k}(\bm{r}')-\alpha_o\frac{\partial f(\widehat{o}_k,\widehat{\sigma}_k)}{\partial \widehat{o}}(\bm{r}')-\frac{\partial g(\widehat{o}_k)}{\partial \widehat{o}}(\bm{r}'),
\label{eq_method_update_o}\\
\widehat{\sigma}_{k+1}&=\widehat{\sigma}_{k}-\alpha_{\sigma}\frac{\partial f(\widehat{o}_k,\widehat{\sigma}_k)}{\partial \widehat{\sigma}},
\label{eq_method_update_sigma}
\end{align}
where $k$ is the index of the iterations, and $\alpha_o$ and $\alpha_{\sigma}$ are parameters for tuning the steps.

We derive the partial derivatives in Eqs.~(\ref{eq_method_update_o}) and (\ref{eq_method_update_sigma}) by using the chain rule~\cite{Kaare2012}.
In Eq.~(\ref{eq_method_update_o}), the partial derivative in the second term on the right side is calculated as
\begin{equation}
\begin{aligned}
\frac{\partial f(\widehat{o},\widehat{\sigma})}{\partial \widehat{o}}(\bm{r}')
&=\frac{\partial \mathcal{A}(\widehat{o}\gamma)}{\partial \widehat{o}}(\bm{r}') \cdot \frac{\partial e}{\partial \mathcal{A}(\widehat{o}\gamma)}(\bm{r}') \cdot \frac{\partial f}{\partial e}(\bm{r}')\\
&=\gamma(\bm{r}')\left[\mathcal{F}^{-1}\bigl(2\mathcal{F}(\widehat{o}\gamma)\mathcal{F} (2 d v^{2} e)\bigr)\right](\bm{r}'),
\end{aligned}
\label{eq_method_partial_f_o}
\end{equation}
where $\mathcal{F}$ and $\mathcal{F}^{-1}$ denote the Fourier transform and the inverse Fourier transform, respectively.
The partial derivative in the third term is written as
\begin{equation}
\begin{aligned}
\frac{\partial g(\widehat{o})}{\partial \widehat{o}}(\bm{r}')
&=\alpha_+ \frac{\partial g_+(\widehat{o})}{\partial\widehat{o}}(\bm{r}')+ \alpha_{\theta} \frac{\partial g_{\theta}(\widehat{o})}{\partial\widehat{o}}(\bm{r}') + \alpha_{\ell_1} \frac{\partial g_{\ell_1}}{\partial\widehat{o}}(\bm{r}')\\
&=2\alpha_+\min(0,\widehat{o}(\bm{r}'))
+2\alpha_{\theta}(\max(\theta,\widehat{o}(\bm{r}'))-\theta)
+\alpha_{\ell_1}.
\end{aligned}
\label{eq_method_partial_g_o}
\end{equation}
Each result of the derivatives in Eqs.~(\ref{eq_method_partial_f_o}) and (\ref{eq_method_partial_g_o}) does not have imaginary parts when $\widehat{o}$ is composed of real numbers, and thus the realness of the updated object is satisfied.
In Eq.~(\ref{eq_method_update_sigma}), the partial derivative in the second term on the right side is calculated as
\begin{equation}
\begin{aligned}
\frac{\partial f(\widehat{o},\widehat{\sigma})}{\partial \widehat{\sigma}}
&=\frac{\partial d(\bm{r})}{\partial \widehat{\sigma}} \cdot \frac{\partial e}{\partial d}(\bm{r}) \cdot \frac{\partial f}{\partial e}(\bm{r})\\
&=\bigintsss \frac{2\widehat{\sigma} |\bm{r}|^2(\sinh(\widehat{\sigma} |\bm{r}|)-\widehat{\sigma}|\bm{r}|\cosh(\widehat{\sigma} |\bm{r}|))}{\sinh^3(\widehat{\sigma} |\bm{r}|)} \left[\mathcal{A}(\widehat{o}\gamma) v^{2} 2e\right](\bm{r})~\mathrm{d}^2\bm{r}.
\end{aligned}
\label{eq_method_partial_f_sigma}
\end{equation}
The updating processes in Eqs.~(\ref{eq_method_update_o}) and (\ref{eq_method_update_sigma}) iterate until the cost function in Eq.~(\ref{eq_method_opt}) converges, and then the limited autocorrelation is extrapolated.

\section{Demonstration and Discussion}

\subsection{Simulation}

We conducted a simulation to quantitatively analyze the proposed method compared with the conventional one.
In this simulation, the pixel count~$N^2$ of the object~$o$ was $128\times 128$, and point sources were randomly located on a $40\times 40$~pixel central area of the object.
The density~$\rho$ of the point sources in the area was varied from 0.005 to 0.03 at intervals of 0.005.
The decay parameter~$\sigma$ was also varied from 0 to 0.14 at intervals of 0.02, where $\sigma=0$ corresponded to the decay-free memory effect.
The reconstruction results at each density~$\rho$ and decay parameter~$\sigma$ were evaluated with the peak signal-to-noise ratio~(PSNR)~\cite{Dong2016, Wang2019}, which was defined as
\begin{equation}
\mathrm{PSNR}=10\log_{10}\frac{\max(o)^2}{\frac{1}{N^2}\int \left(o(\bm{r}')-\widehat{o}(\bm{r}')\right)^2\mathrm{d}^2\bm{r}'}.
\label{eq_psnr}
\end{equation}
As a reference, a reconstruction algorithm without considering the decay function was also employed.
In this conventional approach, the estimated decay parameter~$\widehat{\sigma}$ and the tuning parameter~$\alpha_{\sigma}$ were set to 0 in the updating step of Eq.~(\ref{eq_method_update_sigma}) in the proposed algorithm so that the decay function could be ignored.
We refer to this simplified version as the conventional algorithm.

In both the proposed and conventional algorithms, the object's support~$\gamma$ was a binary square of $64\times 64$~pixels, and the autocorrelation's support~$v$ was a circular hole with a diameter of 5~pixels.
The object was recovered from the decayed autocorrelation with 60,000~iterations in the algorithms.
The tuning parameters in the algorithms were set to $\theta=1.5$, $\alpha_+ = \alpha_{\theta}=5\times 10^{-3}$, $\alpha_{\ell_1}=6\times 10^{-7}$, and $\alpha_o =2\times 10^{-4}$.
$\alpha_{\sigma}$ in the proposed algorithm was set to $2\times 10^{-7}$.
We empirically tuned these parameters by supposing a unimodal potential map of the inverse problem.
This parameter tuning process may be automated by using a grid search to find the parameters that minimize the error function~$f$ in Eq.~(\ref{eq_method_errorfunc}).
The algorithms started from ten combinations of random patterns for the initial~$\widehat{o}$ and random values for the initial~$\widehat{\sigma}$, and one result that achieved the minimal error function~$f$ was chosen~\cite{Bertolotti2012,Katz2014a}.
The calculation time for this reconstruction process was about 20~minutes using MATLAB with an Intel Xeon Gold 6254 processor running at 3.10~GHz and equipped with 376~GB of RAM.
This calculation time may be reduced by using parallel processing with a GPU and accelerating the gradient step by employing algorithms that use momentum~\cite{Ruder2016}.

\begin{figure}[t!]
\begin{center}
		\subfigure[]{\label{fig_simulation_psnr_proposed}\includegraphics[scale=1]{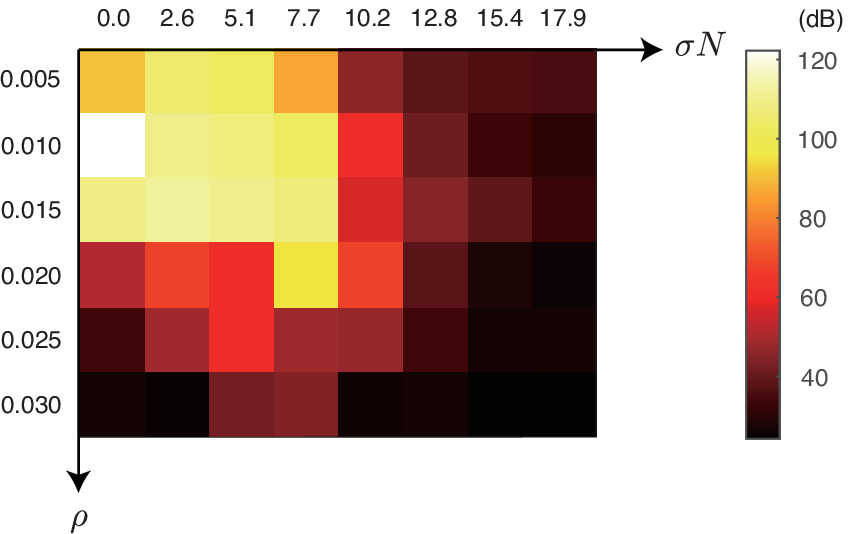}}
		\subfigure[]{\label{fig_simulation_psnr_conventional}\includegraphics[scale=1]{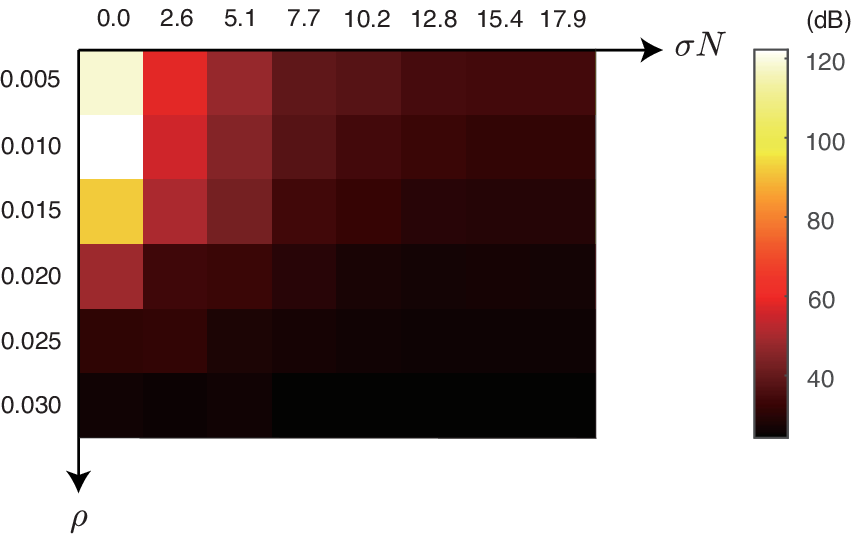}}\end{center}
    \caption{PSNRs of the object estimation depending on the normalized decay parameter~$\sigma N$ and the density~$\rho$ by the \subref{fig_simulation_psnr_proposed}~proposed and \subref{fig_simulation_psnr_conventional}~conventional algorithms.}
\label{fig_simulation_psnr}
\end{figure}

\begin{figure}[t!]
    \centering
    \includegraphics[scale=1]{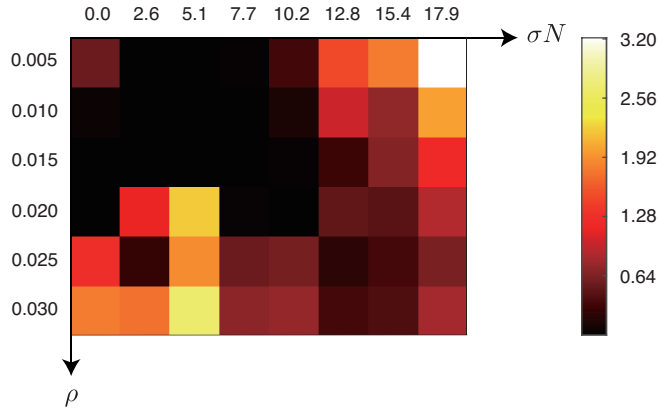}
    \caption{RMSEs of the normalized estimated decay parameter~$\widehat{\sigma} N$ depending on the normalized decay parameter~$\sigma N$ and the density~$\rho$ by the proposed algorithm.}
    \label{fig_simulation_sigma}
\end{figure}

The PSNRs depending on the normalized decay parameter~$\sigma N$ and the density~$\rho$ are shown in Fig.~\ref{fig_simulation_psnr}, where each PSNR was the average value calculated from ten randomly generated objects.
Here the normalization of the decay parameter was introduced to compare the simulation results and the experimental results below, where the pixel count~$N$ was different from that in the simulation.
Under most conditions, the proposed algorithm in Fig.~\ref{fig_simulation_psnr_proposed} achieved higher PSNRs compared with the conventional one in Fig.~\ref{fig_simulation_psnr_conventional}.
These results showed the advantage of the proposed algorithm over the conventional one.
As exceptions, the performance of the conventional algorithm was better when the decay parameter~$\sigma$ was 0.
This was because the conventional algorithm assumes a decay-free memory effect.
In both the proposed and conventional algorithms, the PSNRs were low when the object was dense because such objects conflicted with the sparsity regularization in Eq.~(\ref{eq_method_spa}) and the autocorrelation's contrast of such objects degraded.
The PSNRs in the proposed algorithm decreased when the density~$\rho$ was too low or the decay~$\sigma$ was too large, as shown in Fig.~\ref{fig_simulation_psnr_proposed}.
In the former case, sampling points for the decay of the autocorrelation were too sparse and insufficient to estimate the decay because the autocorrelation of such objects was sparse.
The latter case was an under-determinant case because the width of the autocorrelation was too small to estimate the object.
If we define successful reconstructions as those have a PSNR of 70~dB in Fig.~\ref{fig_simulation_psnr}, the number of conditions achieving this PSNR with the proposed method was 13, and that with the conventional method was 3.
In this case, the proposed method increased the number of succesful conditions to 4.3 times that of the conventional method.

In the proposed algorithm, the decay parameter~$\widehat{\sigma}$ was estimated simultaneously with the reconstruction of the object~$o$.
The root-mean-square errors~(RMSEs) between the normalized original and estimated decay parameters under each condition in Fig.~\ref{fig_simulation_psnr_proposed} are shown in Fig.~\ref{fig_simulation_sigma}.
By comparing these two figures, the PSNR of the object estimation was high when the RMSE of the decay parameter estimation was low.
This result shows that the estimations of the object and the decay parameter worked together in the proposed algorithm.
Therefore, the object reconstruction aborted if the estimation of the decay parameter failed, and vice versa.

\subsection{Experiment}

\begin{figure}[t]
    \centering
    \includegraphics[scale=1]{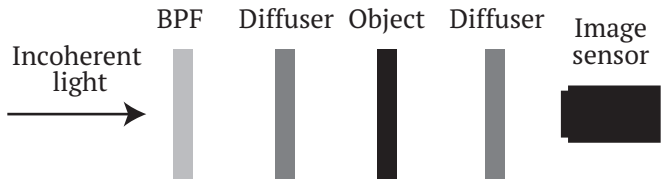}
    \caption{Optical setup of the experimental demonstration.
    BPF:~Bandpass filter.}
    \label{fig_experiment_setup}
\end{figure}

\begin{figure}[t!]
\begin{center}
		\subfigure[]{\label{fig_experiment_capimg}\includegraphics[width=5.0cm]{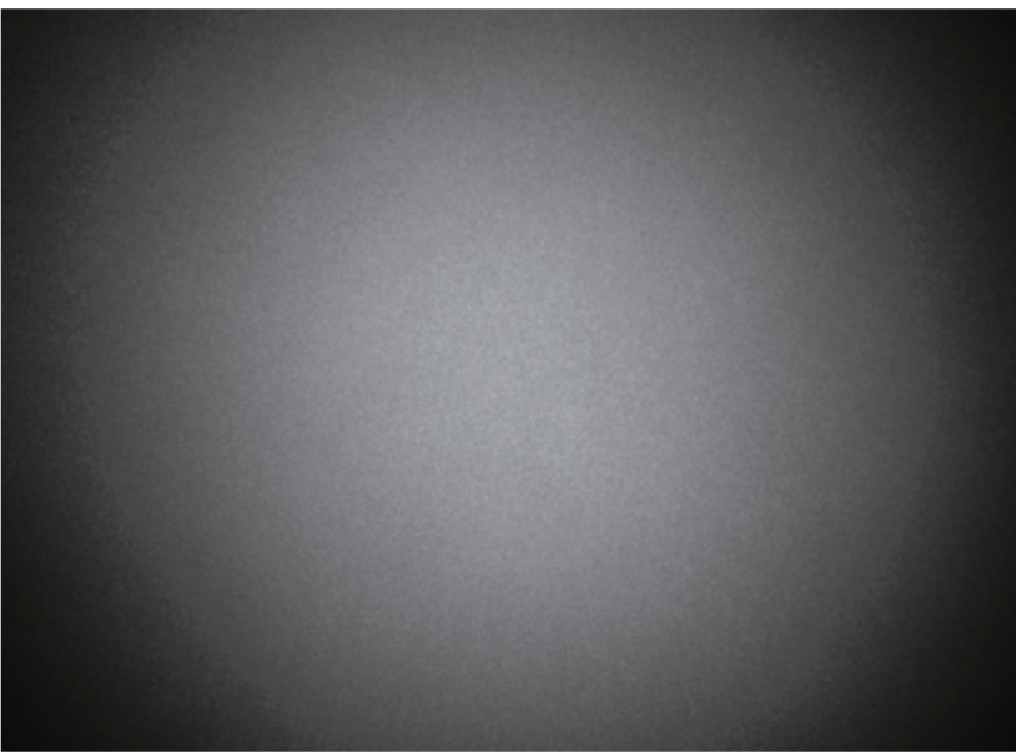}}
		\subfigure[]{\label{fig_experiment_auto}\includegraphics[width=5.0cm]{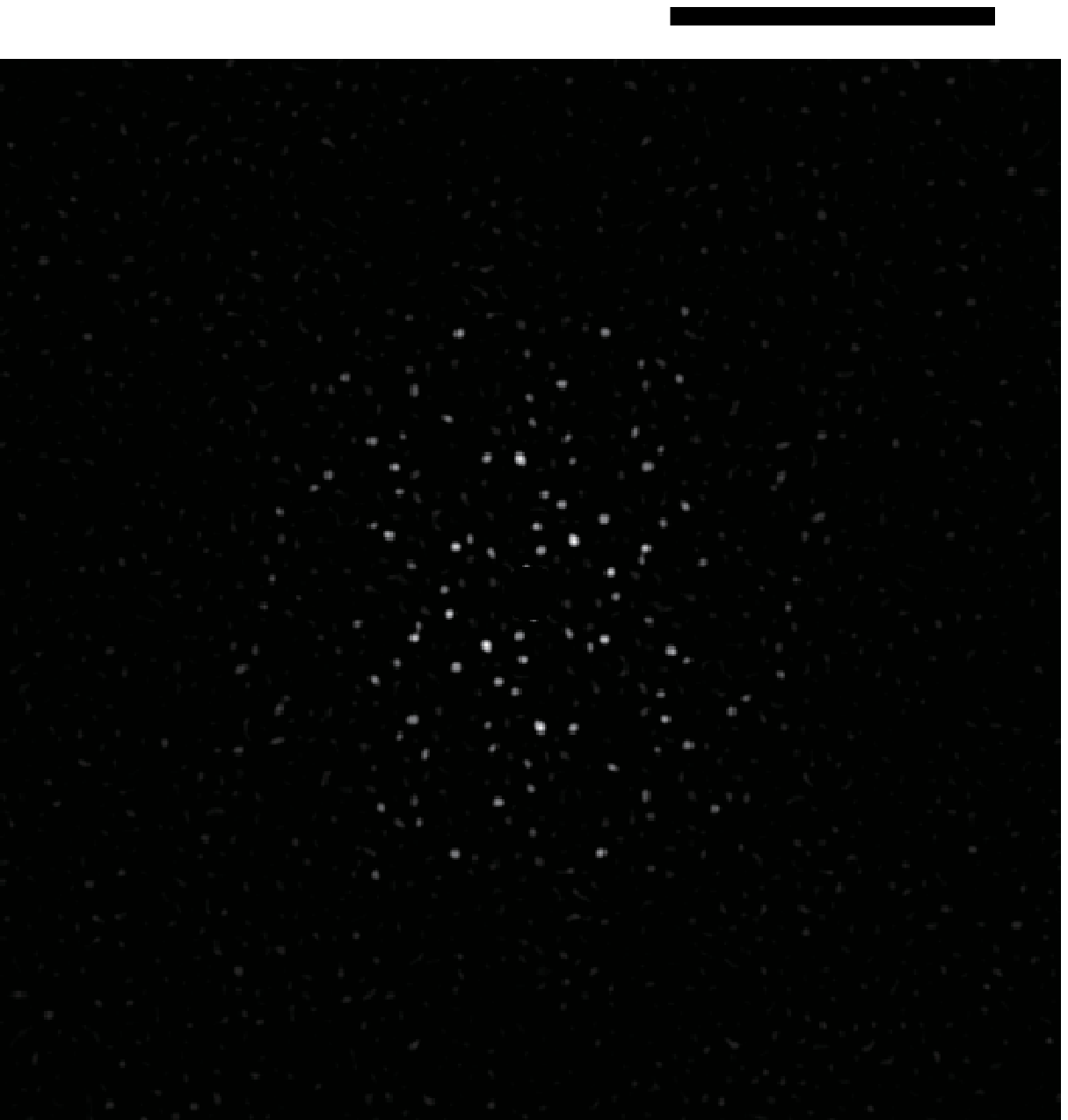}} 
		\\
		\subfigure[]{\label{fig_experiment_object}\includegraphics[width=5.0cm]{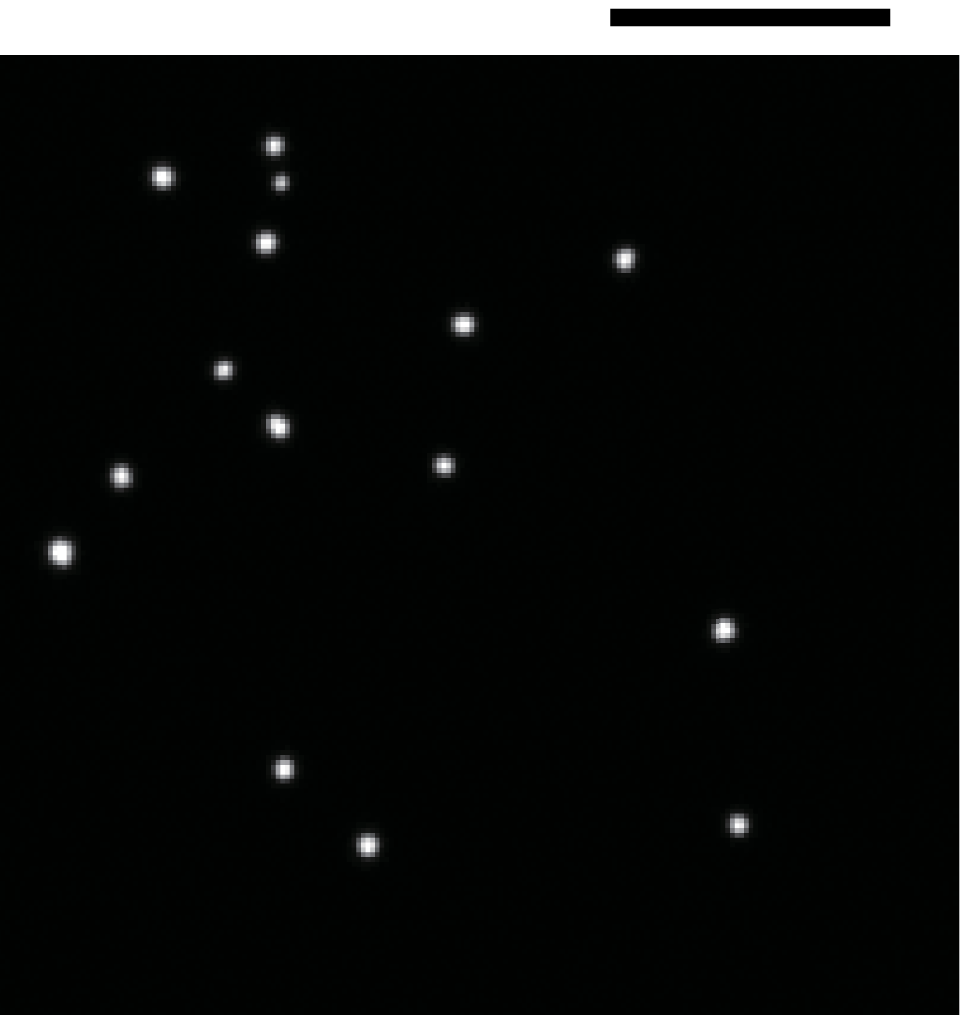}}
		\subfigure[]{\label{fig_experiment_reconstruction_proposed}\includegraphics[width=5.0cm]{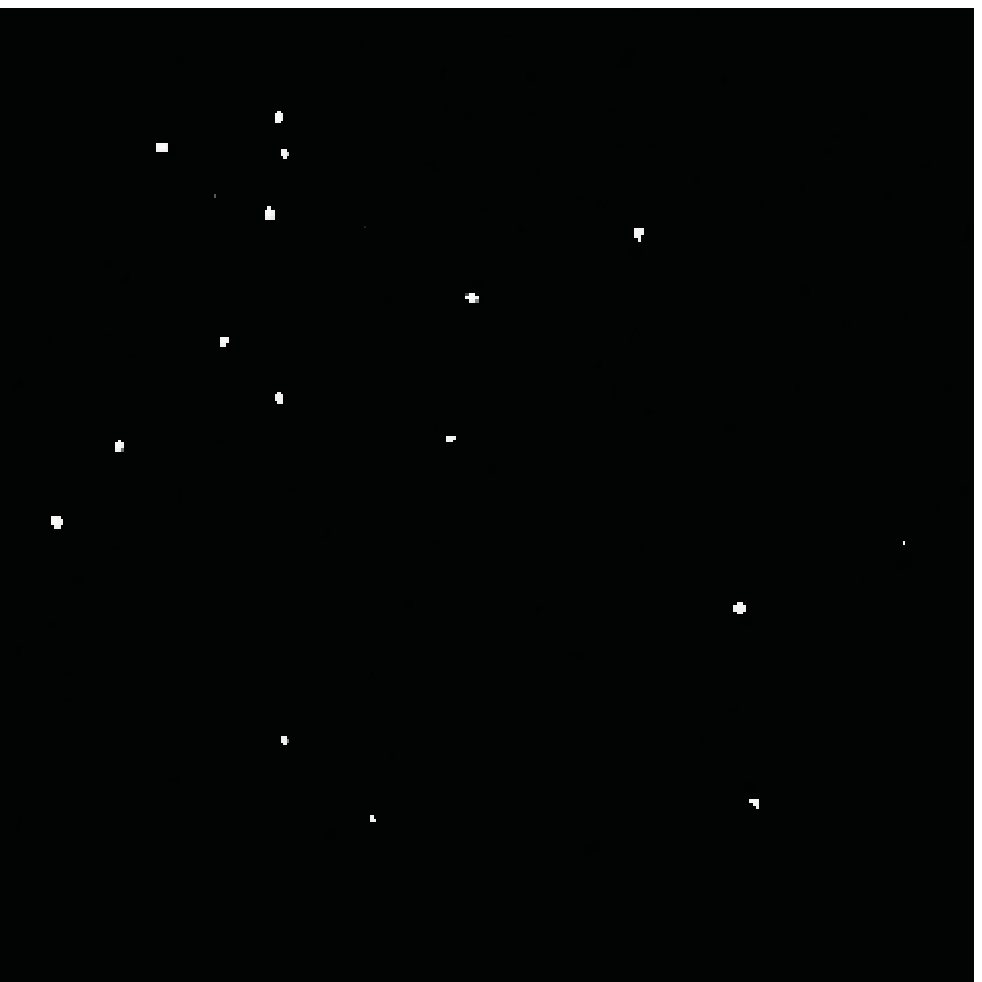}}
		\subfigure[]{\label{fig_experiment_reconstruction_conventional}\includegraphics[width=5.0cm]{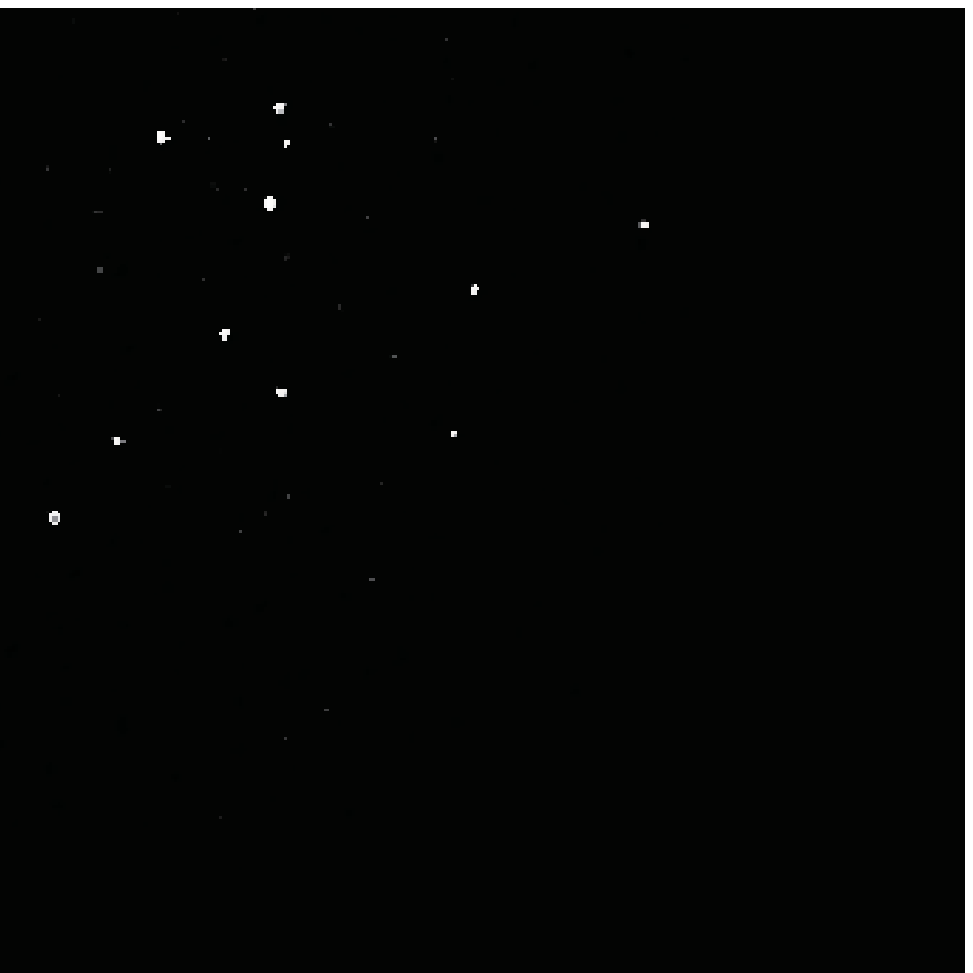}} \\
		\subfigure[]{\label{fig_experiment_original_autocor}\includegraphics[width=5.0cm]{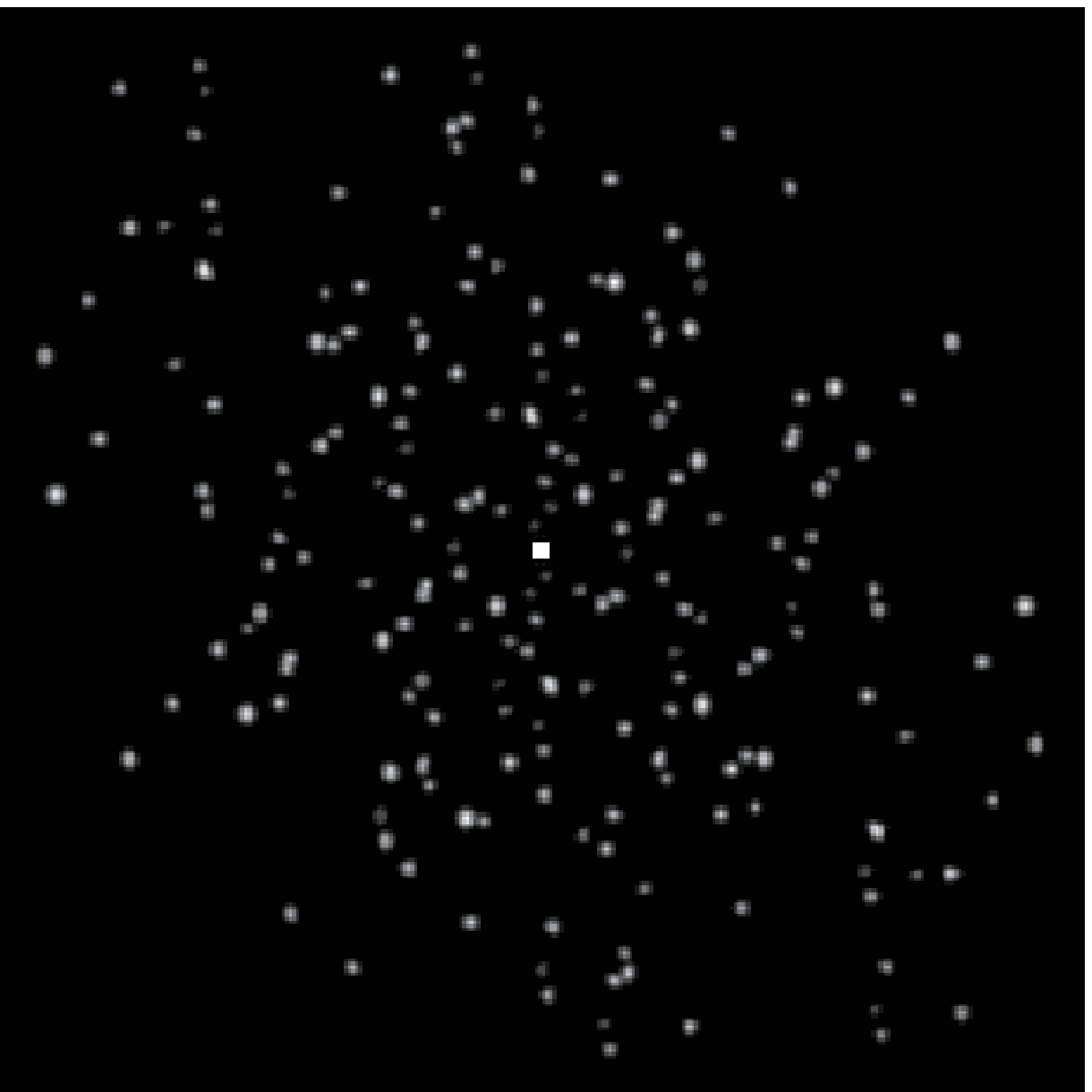}}
		\subfigure[]{\label{fig_experiment_reconstruction_autocor_proposed}\includegraphics[width=5.0cm]{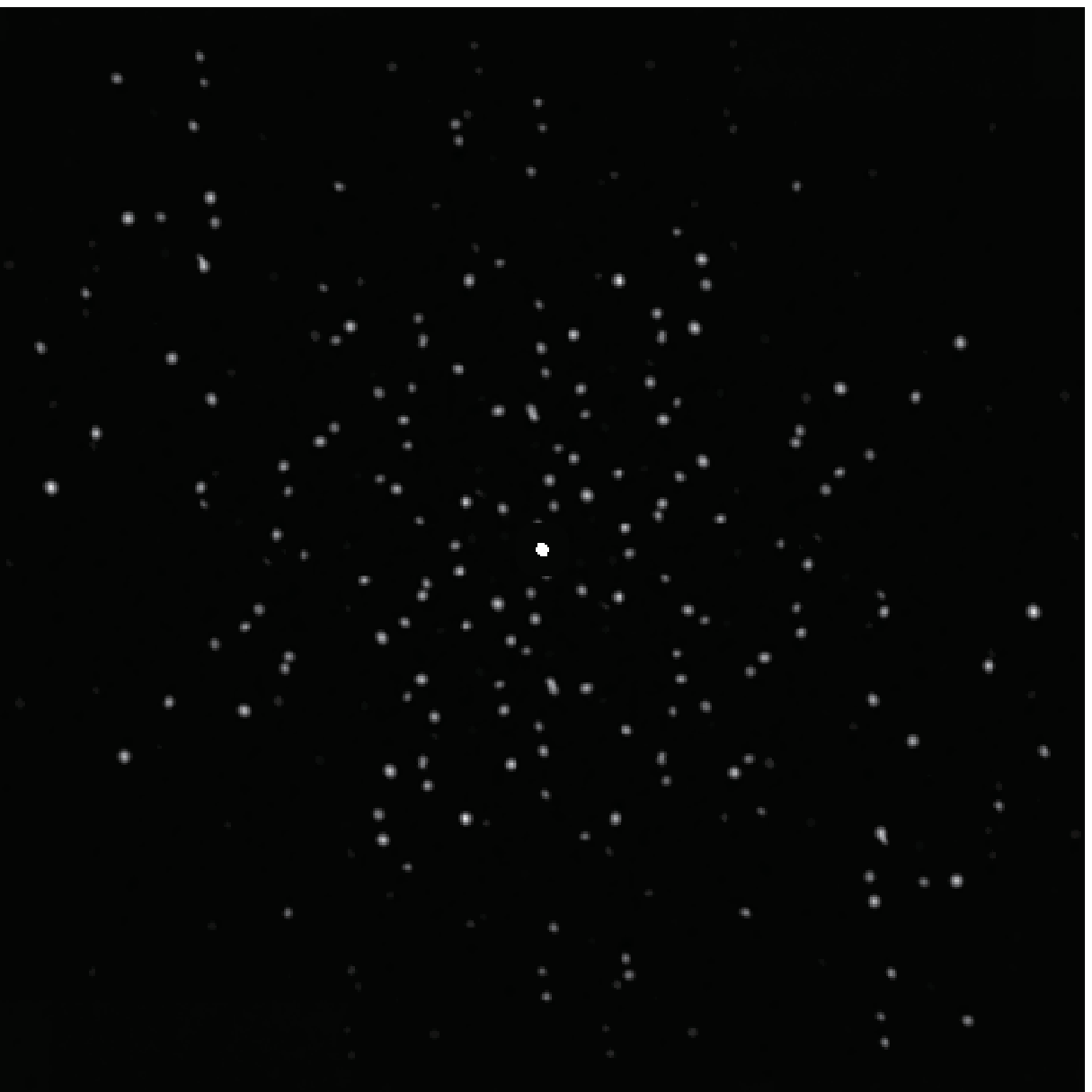}}
		\subfigure[]{\label{fig_experiment_reconstruction_autocor_conventional}\includegraphics[width=5.0cm]{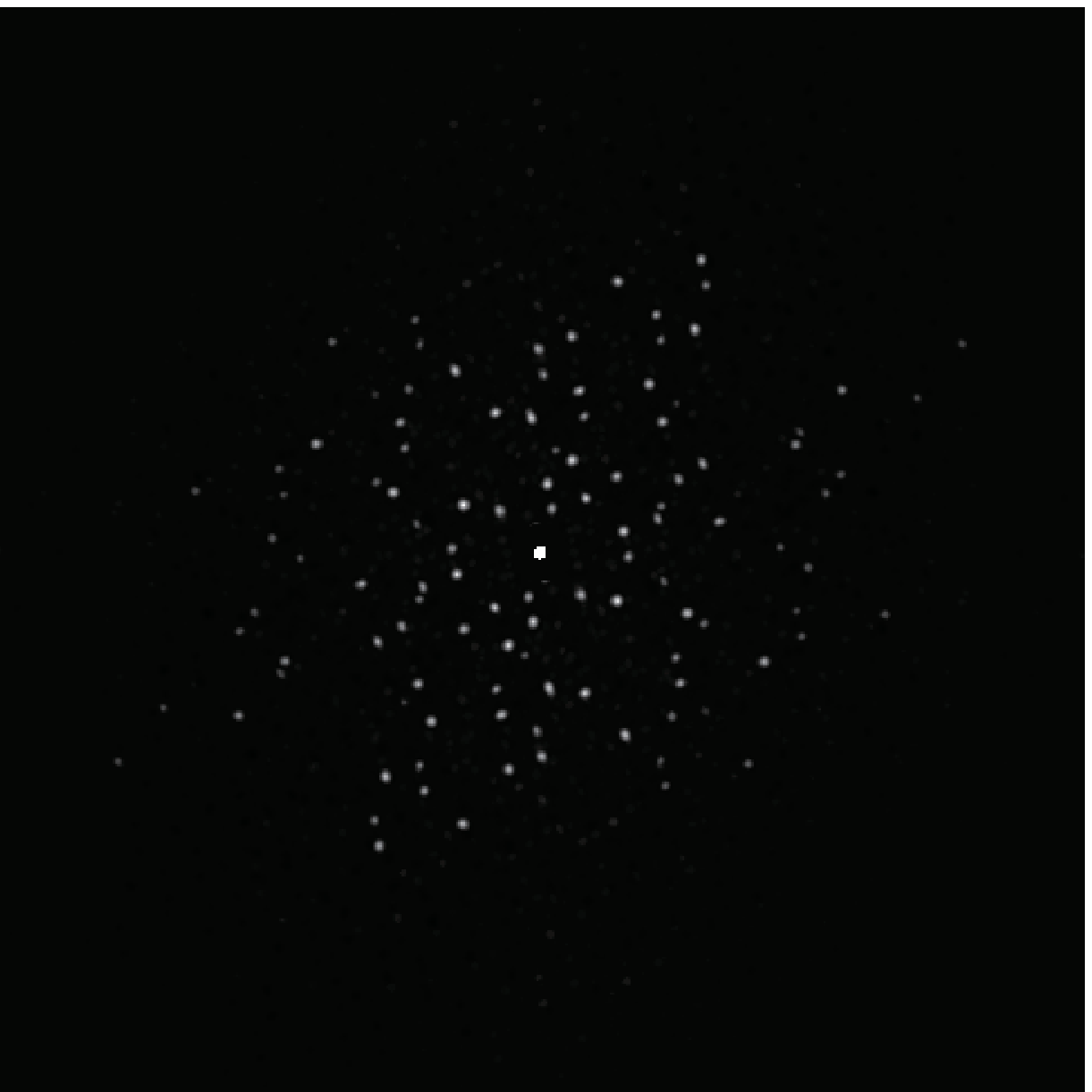}}
\end{center}
\caption{Experimental results.
    \subref{fig_experiment_capimg}~The captured image and \subref{fig_experiment_auto}~its autocorrelation with the support, where the scale bar is 2~mm on the sensor plane.
    \subref{fig_experiment_object}~The object image, where the scale bar is 2~mm on the object plane.
    The reconstructed images obtained by \subref{fig_experiment_reconstruction_proposed}~the proposed algorithm and \subref{fig_experiment_reconstruction_conventional}~the conventional algorithm.
    \subref{fig_experiment_original_autocor}-\subref{fig_experiment_reconstruction_autocor_conventional}~The autocorrelations of \subref{fig_experiment_object}-\subref{fig_experiment_reconstruction_conventional}, respectively.
}
\label{fig_experiment_reconstruction}
\end{figure}

We experimentally demonstrated the proposed method with the optical setup shown in Fig.~\ref{fig_experiment_setup}.
Collimated light from a white light-emitting diode~(LED, XLamp CXA2520 manufactured by CREE) passed through a bandpass filter~(BPF, HMZ0530 manufactured by ASAHI SPECTRA, central wavelength:~530~nm, bandwidth:~10~nm) to implement spatially incoherent monochromatic illumination.
A diffuser~(KHYP1-12 manufactured by Optical Solutions, circular diffusion angle:~5$^\circ$) was illuminated by the spatially incoherent light.
An object, which was a piece of aluminum foil with fifteen holes, as shown in Fig.~\ref{fig_experiment_object}, was illuminated by the diffused light.
Light passing through the object was captured by an image sensor~(DMK38UX253 manufactured by The Imaging Source, pixel count:~$3000\times4096$, pixel pitch:~3.45~\textmu m) through another diffuser~(KHYP1-12 manufactured by Optical Solutions, circular diffusion angle:~5$^\circ$) without any imaging optics. 
The decay parameter~$\sigma$ is variable depending on the distance between the object and the second diffuser, and $\sigma$ becomes larger when the distance is shorter.
In the experiment, the distance between the object and the second diffuser was 40~mm, and the distance between the second diffuser and the image sensor was 27~mm, respectively.

The captured image is shown in Fig.~\ref{fig_experiment_capimg}, where the point sources are not recognizable.
To compensate for the shading effect, the captured image was divided by a blurred version of the captured speckle image with low-pass filtering.
Then, the autocorrelation of the compensated image was calculated.
The central region of $2,400\times 2,400$~pixels of the autocorrelation was clipped after subtracting the background noise~$\widetilde{c}$ in Eq.~(\ref{eq_method_forwardmodel_2}), which was determined with the maximal pixel value outside of the clipped area.
Then, the pixel count~$N^2$ was resized to $1,200\times 1,200$~pixels to reduce the computational cost.
The autocorrelation result with the support~$v(\bm{r})$ is shown in Fig.~\ref{fig_experiment_auto}.
The proposed and conventional algorithms were applied to the resultant image, as mentioned in the simulation section above.
In the experiment, the object's support~$\gamma$ was a binary square of $600\times 600$~pixels, and the autocorrelation's support~$v$ was a circular hole with a diameter of 50~pixels.
The tuning parameters were set to $\theta=3.2\times 10^{-2}$, $\alpha_+ = \alpha_{\theta}=1\times 10^{-2}$, $\alpha_{\ell_1}=1\times 10^{-6}$, and $\alpha_o =1\times 10^{-4}$, in the both algorithms, and $\alpha_{\sigma}=1\times 10^{-7}$ in the proposed algorithm.
The other conditions were the same as the above simulation.

After estimating the object~$\widehat{o}$ and the decay parameter~$\widehat{\sigma}$ in the proposed and conventional algorithms, a denoising process was additionally applied to the results of both algorithms by using a modified version of the conventional algorithm, where the decay parameter~$\widehat{\sigma}$ was not updated because $\alpha_{\sigma}=0$.
In the modified conventional algorithm, the penalty term in Eq.~(\ref{eq_method_spa}) was replaced with $g_{\ell_1}(\widehat{o})=\|\max(0,\widehat{o}(\{\bm{r}'|~\widehat{o}(\bm{r}')<\theta'\})\|_1$ to remove small noise defined by $\theta'$.
The number of iterations in this denoising process was 30,000.

\begin{figure}[t]
    \centering
    \includegraphics[scale=1]{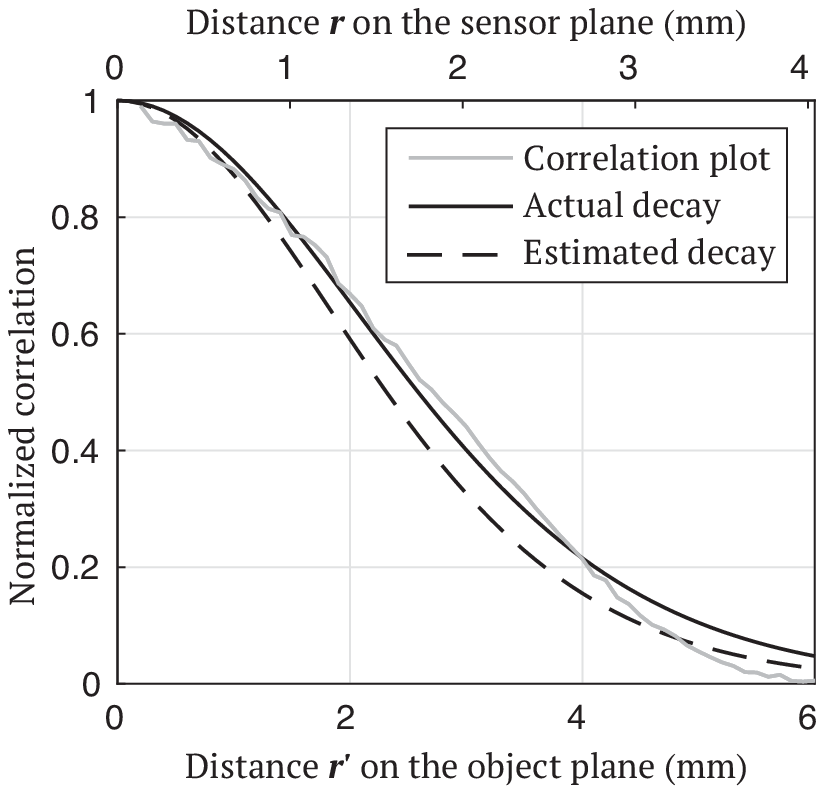}
    \caption{Comparison between the estimated decay function and the actual one, which was calculated by fitting to the experimentally observed correlations.}
    \label{fig_experiment_sigma}
\end{figure}

The reconstruction results obtained by the proposed and the conventional algorithms are shown in Figs.~\ref{fig_experiment_reconstruction_proposed} and \ref{fig_experiment_reconstruction_conventional}, respectively.
By comparing the object image in Fig.~\ref{fig_experiment_object} and these results, the proposed algorithm recovered point sources in a larger area than the conventional one, where the point sources on the lower part disappeared.
The errors calculated in Eq.~(\ref{eq_method_errorfunc}) for the results obtained by the proposed and conventional algorithms were $3.92\times 10^{-1}$ and $4.39\times 10^{-1}$, respectively.
Therefore, the proposed algorithm outperformed the conventional one.

Furthermore, the autocorrelations of the object image in Fig.~\ref{fig_experiment_object} and the reconstructed images in Figs.~\ref{fig_experiment_reconstruction_proposed} and \ref{fig_experiment_reconstruction_conventional} are shown in Figs.~\ref{fig_experiment_original_autocor}-\ref{fig_experiment_reconstruction_autocor_conventional}, respectively.
By comparing these autocorrelations, the autocorrelation estimated by the proposed algorithm was broader than that by the conventional one.
This result verified our concept of speckle-correlation imaging with the extended field-of-view by extrapolating the limited autocorrelation.
In addition, both the proposed and conventional algorithms interpolated the central peak of the autocorrelation removed by the autocorrelation support shown in Fig.~\ref{fig_experiment_auto}.

In the experiment, the proposed algorithm estimated the normalized decay parameter~$\widehat{\sigma} N$ as $7.93$, and the estimated decay function~$d(\widehat{\sigma},\bm{r})$ is shown in Fig.~\ref{fig_experiment_sigma}.
The actual correlations~$p(\bm{r})$ of the PSFs depending on the lateral position~$\bm{r}$ were experimentally measured by scanning a point source with intervals of 0.1~mm.
The actual normalized decay parameter~$\sigma_{\text{act}} N$ was calculated to be $7.09$ from the experimentally observed correlations by solving $\text{arg~min}_{\sigma_{\text{act}},a}\|(d(\sigma_{\text{act}},\bm{r})-ap(\bm{r}))v(\bm{r})\|_2^2$ based on the least-squares method.
The actual decay function~$d(\sigma_{\text{act}},\bm{r})$ and the correlation plot~$ap(\bm{r})v(\bm{r})$ are also shown in Fig.~\ref{fig_experiment_sigma}.
In this case, the RMSE of the normalized estimated decay parameter~$\widehat{\sigma} N$ was 0.84.
The density~$\rho$ of the object in the experiment was $0.013$.
The normalized decay parameter~$\sigma N$ and density~$\rho$ in the experiment corresponded to the condition where the proposed method was applicable but the conventional method was not applicable, as shown in Fig.~\ref{fig_simulation_psnr}.
In addition, the experimental RMSE was slightly higher than the simulated one at the normalized decay parameter and the density shown in Fig.~\ref{fig_simulation_sigma}, and this RMSE was permissible for the object reconstruction.
From these results of the reconstructed object~$\widehat{o}$ and the estimated decay parameter~$\widehat{\sigma}$, it was verified that the proposed method extended the field of view under the limited memory effect.

\section{Conclusion}
We presented and demonstrated a method for extending the field of view of single-shot speckle-correlation imaging under a limited memory effect.
We modeled the speckle correlation taking into account such a situation.
Then, we derived the inverse process to the object and the decay parameter of the correlation from the speckle correlation.
The object and the decay parameter were simultaneously estimated based on the gradient descent algorithm, and as a result, the limited autocorrelation was extrapolated.
We verified the proposed method in both a simulation and an experiment.
The simulation result showed that the proposed algorithm outperformed the conventional algorithm when the memory effect was limited.
In the experiment, the proposed algorithm recovered a larger field of view compared with the conventional algorithm, and its estimated decay parameter agreed with the actually measured one.

The proposed method is readily applied to conventional speckle-correlation imaging without any optical modification.
Therefore, it enhances the practicality of imaging techniques for seeing through scattering media with non-invasiveness and a minimal lensless setup.
In this study, we assumed spatially sparse point sources as objects for the proof-of-concept, and this assumption is acceptable in several applications, such as astronomical observation and super-resolution fluorescence microscopy~\cite{Betzig2006, Davies2012, Tehrani2015, Wang2021}.
The sparsity assumption in our method may be mitigated by using phase retrieval based on compressive sensing and deep learning~\cite{Baraniuk2007,Horisaki2014, Nishizaki2020, Metzler2020}.
It is also extendable to multidimensional speckle-correlation imaging~\cite{Okamoto2019,Horisaki2019,Ehira2021}.
Therefore, the proposed method will contribute to imaging applications in various fields, such as biomedicine, astronomy, and security.



\begin{thebibliography}{10}
\newcommand{\enquote}[1]{``#1''}

\bibitem{Ntziachristos2010}
V.~Ntziachristos, \enquote{Going deeper than microscopy: The optical imaging
  frontier in biology,} {\protect\JournalTitle{Nature Methods}} \textbf{7},
  603--614 (2010).

\bibitem{Ji2017}
N.~Ji, \enquote{Adaptive optical fluorescence microscopy,}
  {\protect\JournalTitle{Nature Methods}} \textbf{14}, 374--380 (2017).

\bibitem{Davies2012}
R.~Davies and M.~Kasper, \enquote{Adaptive optics for astronomy,}
  {\protect\JournalTitle{Annual Review of Astronomy and Astrophysics}}
  \textbf{50}, 305--351 (2012).

\bibitem{Watnik2018}
A.~T. Watnik and D.~F. Gardner, \enquote{Wavefront sensing in deep turbulence,}
  {\protect\JournalTitle{Opt. Photon. News}} \textbf{29}, 38--45 (2018).

\bibitem{Mosk2012}
A.~P. Mosk, A.~Lagendijk, G.~Lerosey, and M.~Fink, \enquote{Controlling waves
  in space and time for imaging and focusing in complex media,}
  {\protect\JournalTitle{Nature Photonics}} \textbf{6}, 283--292 (2012).

\bibitem{Horstmeyer2015}
R.~Horstmeyer, H.~Ruan, and C.~Yang, \enquote{Guidestar-assisted
  wavefront-shaping methods for focusing light into biological tissue,}
  {\protect\JournalTitle{Nature Photonics}} \textbf{9}, 563--571 (2015).

\bibitem{Yoon2020}
S.~Yoon, M.~Kim, M.~Jang, Y.~Choi, W.~Choi, S.~Kang, and W.~Choi, \enquote{Deep
  optical imaging within complex scattering media,}
  {\protect\JournalTitle{Nature Reviews Physics}} \textbf{2}, 141--158 (2020).

\bibitem{Vellekoop2007}
I.~M. Vellekoop and A.~P. Mosk, \enquote{Focusing coherent light through opaque
  strongly scattering media,} {\protect\JournalTitle{Optics Letters}}
  \textbf{32}, 2309--2311 (2007).

\bibitem{Vellekoop2010a}
I.~M. Vellekoop, A.~Lagendijk, and A.~P. Mosk, \enquote{Exploiting disorder for
  perfect focusing,} {\protect\JournalTitle{Nature Photonics}} \textbf{4},
  320--322 (2010).

\bibitem{Katz2011}
O.~Katz, E.~Small, Y.~Bromberg, and Y.~Silberberg, \enquote{Focusing and
  compression of ultrashort pulses through scattering media,}
  {\protect\JournalTitle{Nature Photonics}} \textbf{5}, 372--377 (2011).

\bibitem{Popoff2010a}
S.~Popoff, G.~Lerosey, M.~Fink, A.~C. Boccara, and S.~Gigan, \enquote{Image
  transmission through an opaque material,} {\protect\JournalTitle{Nature
  Communications}} \textbf{1} (2010).

\bibitem{Liutkus2014}
A.~Liutkus, D.~Martina, S.~Popoff, G.~Chardon, O.~Katz, G.~Lerosey, S.~Gigan,
  L.~Daudet, and I.~Carron, \enquote{Imaging with nature: Compressive imaging
  using a multiply scattering medium,} {\protect\JournalTitle{Scientific
  Reports}} \textbf{4}, 5552 (2014).

\bibitem{Horisaki2016}
R.~Horisaki, R.~Takagi, and J.~Tanida, \enquote{Learning-based imaging through
  scattering media,} {\protect\JournalTitle{Opt. Express}} \textbf{24},
  13738--13743 (2016).

\bibitem{Bertolotti2012}
J.~Bertolotti, E.~G.~V. Putten, C.~Blum, A.~Lagendijk, W.~L. Vos, and A.~P.
  Mosk, \enquote{Non-invasive imaging through opaque scattering layers,}
  {\protect\JournalTitle{Nature}} \textbf{491}, 232--234 (2012).

\bibitem{Katz2014a}
O.~Katz, P.~Heidmann, M.~Fink, and S.~Gigan, \enquote{Non-invasive single-shot
  imaging through scattering layers and around corners via speckle
  correlations,} {\protect\JournalTitle{Nature Photonics}} \textbf{8}, 784--790
  (2014).

\bibitem{Fienup1982}
J.~R. Fienup, \enquote{Phase retrieval algorithms: a comparison,}
  {\protect\JournalTitle{Appl. Opt.}} \textbf{21}, 2758--2769 (1982).

\bibitem{Feng1988}
S.~Feng, C.~Kane, P.~A. Lee, and A.~D. Stone, \enquote{Correlations and
  fluctuations of coherent wave transmission through disordered media,}
  {\protect\JournalTitle{Physical Review Letters}} \textbf{61}, 834--837
  (1988).

\bibitem{Issac1988}
I.~Freund, M.~Rosenbluh, and S.~Feng, \enquote{Memory effects in propagation of
  optical waves through disordered media,} {\protect\JournalTitle{Physical
  Review Letters}} \textbf{61}, 2328--2331 (1988).

\bibitem{Fienup2013}
J.~R. Fienup, \enquote{Phase retrieval algorithms: a personal tour,}
  {\protect\JournalTitle{Appl. Opt.}} \textbf{52}, 45--56 (2013).

\bibitem{Okamoto2019}
Y.~Okamoto, R.~Horisaki, and J.~Tanida, \enquote{Noninvasive three-dimensional
  imaging through scattering media by three-dimensional speckle correlation,}
  {\protect\JournalTitle{Opt. Lett.}} \textbf{44}, 2526--2529 (2019).

\bibitem{Horisaki2019}
R.~Horisaki, Y.~Okamoto, and J.~Tanida, \enquote{Single-shot noninvasive
  three-dimensional imaging through scattering media,}
  {\protect\JournalTitle{Opt. Lett.}} \textbf{44}, 4032--4035 (2019).

\bibitem{Ehira2021}
K.~Ehira, R.~Horisaki, Y.~Nishizaki, M.~Naruse, and J.~Tanida,
  \enquote{Spectral speckle-correlation imaging,}
  {\protect\JournalTitle{Applied Optics}} \textbf{60}, 2388--2392 (2021).

\bibitem{Li2019a}
G.~Li, W.~Yang, H.~Wang, and G.~Situ, \enquote{Image transmission through
  scattering media using ptychographic iterative engine,}
  {\protect\JournalTitle{Applied Sciences (Switzerland)}} \textbf{9}, 849
  (2019).

\bibitem{Rosenfeld2021}
M.~Rosenfeld, G.~Weinberg, D.~Doktofsky, Y.~Li, L.~Tian, and O.~Katz,
  \enquote{Acousto-optic ptychography,} {\protect\JournalTitle{Optica}}
  \textbf{8}, 936--943 (2021).

\bibitem{Wang2019}
X.~Wang, X.~Jin, J.~Li, X.~Lian, X.~Ji, and Q.~Dai,
  \enquote{Prior-information-free single-shot scattering imaging beyond the
  memory effect,} {\protect\JournalTitle{Optics Letters}} \textbf{44},
  1423--1426 (2019).

\bibitem{Alterman2021}
M.~Alterman, C.~Bar, I.~Gkioulekas, and A.~Levin, \enquote{Imaging with local
  speckle intensity correlations: Theory and practice,}
  {\protect\JournalTitle{ACM Transactions on Graphics}} \textbf{40}, 1--22
  (2021).

\bibitem{Schot2015}
S.~Schott, J.~Bertolotti, J.-F. L\'{e}ger, L.~Bourdieu, and S.~Gigan,
  \enquote{Characterization of the angular memory effect of scattered light in
  biological tissues,} {\protect\JournalTitle{Opt. Express}} \textbf{23},
  13505--13516 (2015).

\bibitem{Hofer2018}
M.~Hofer, C.~Soeller, S.~Brasselet, and J.~Bertolotti, \enquote{Wide field
  fluorescence epi-microscopy behind a scattering medium enabled by speckle
  correlations,} {\protect\JournalTitle{Opt. Express}} \textbf{26}, 9866--9881
  (2018).

\bibitem{Betzig2006}
E.~Betzig, G.~H. Patterson, R.~Sougrat, O.~W. Lindwasser, S.~Olenych, J.~S.
  Bonifacino, M.~W. Davidson, J.~Lippincott-Schwartz, and H.~F. Hess,
  \enquote{Imaging intracellular fluorescent proteins at nanometer resolution,}
  {\protect\JournalTitle{Science}} \textbf{313}, 1642--1645 (2006).

\bibitem{Tehrani2015}
K.~F. Tehrani, J.~Xu, Y.~Zhang, P.~Shen, and P.~Kner, \enquote{Adaptive optics
  stochastic optical reconstruction microscopy ({AO-STORM}) using a genetic
  algorithm,} {\protect\JournalTitle{Opt. Express}} \textbf{23}, 13677--13692
  (2015).

\bibitem{Wang2021}
D.~Wang, S.~K. Sahoo, X.~Zhu, G.~Adamo, and C.~Dang, \enquote{Non-invasive
  super-resolution imaging through dynamic scattering media,}
  {\protect\JournalTitle{Nature Communications}} \textbf{12}, 3150 (2021).

\bibitem{Kaare2012}
K.~B. Petersen and M.~S. Pedersen, \enquote{The matrix cookbook,}  (2012).

\bibitem{Dong2016}
C.~Dong, C.~C. Loy, K.~He, and X.~Tang, \enquote{Image super-resolution using
  deep convolutional networks,} {\protect\JournalTitle{IEEE Transactions on
  Pattern Analysis and Machine Intelligence}} \textbf{38}, 295--307 (2016).

\bibitem{Ruder2016}
S.~Ruder, \enquote{An overview of gradient descent optimization algorithms,}
  {\protect\JournalTitle{CoRR}} \textbf{abs/1609.04747} (2016).

\bibitem{Baraniuk2007}
R.~G. Baraniuk, \enquote{Compressive sensing,} {\protect\JournalTitle{IEEE
  Signal Processing Magazine}} \textbf{24}, 118--121 (2007).

\bibitem{Horisaki2014}
R.~Horisaki, Y.~Ogura, M.~Aino, and J.~Tanida, \enquote{Single-shot phase
  imaging with a coded aperture,} {\protect\JournalTitle{Optics letters}}
  \textbf{39}, 6466--6469 (2014).

\bibitem{Nishizaki2020}
Y.~Nishizaki, R.~Horisaki, K.~Kitaguchi, M.~Saito, and J.~Tanida,
  \enquote{Analysis of non-iterative phase retrieval based on machine
  learning,} {\protect\JournalTitle{Optical Review}} \textbf{27}, 136--141
  (2020).

\bibitem{Metzler2020}
C.~A. Metzler, F.~Heide, P.~Rangarajan, M.~M. Balaji, A.~Viswanath,
  A.~Veeraraghavan, and R.~G. Baraniuk, \enquote{Deep-inverse correlography:
  towards real-time high-resolution non-line-of-sight imaging,}
  {\protect\JournalTitle{Optica}} \textbf{7}, 63--71 (2020).

\end{thebibliography}

\end{document}